# Context-based Broadcast Acknowledgement for Enhanced Reliability of Cooperative V2X Messages

Baldomero Coll-Perales, Gokulnath Thandavarayan, Miguel Sepulcre and Javier Gozalvez

*Abstract*— Most V2X applications/services are supported by the continuous exchange of broadcast messages. One of the main challenges is to increase the reliability of broadcast transmissions that lack of mechanisms to assure the correct delivery of the messages. To address this issue, one option is the use of acknowledgments. However, this option has scalability issues when applied to broadcast transmissions because multiple vehicles can transmit acknowledgments simultaneously. To control scalability while addressing reliability of broadcast messages, this paper proposes and evaluates a context-based broadcast acknowledgement mechanism where the transmitting vehicles selectively request the acknowledgment of specific/critical broadcast messages, and performs retransmissions if they are not correctly received. In addition, the V2X applications/services identify the situations/conditions that trigger the execution of the broadcast acknowledgment mechanism, and the receiver(s) that should acknowledge the broadcast messages. The paper evaluates the performance of the context-based broadcast acknowledgment mechanism for a Collective Perception Service. The obtained results show the proposed mechanism can contribute to improve the awareness of crossing pedestrians at intersections by increasing the reliability in the exchange of CPM messages between vehicles approaching the intersection. This solution is being discussed under IEEE 802.11bd, and thus can be relevant for the standardization process.

## I. INTRODUCTION

Connected vehicles rely on V2X (Vehicle to Everything) communication technologies to expand their capabilities by utilizing contextual information beyond the local environment they can perceive. Based on the Car 2 Car Communication Consortium (C2C-CC) roadmap published in [1], the first phase in the deployment of V2X communications (a.k.a. Day1) will include Cooperative Awareness (CA) and Decentralized Environment Notification (DEN) services to disseminate the vehicles' status information (location, speed, acceleration, heading direction, etc.) as well as the occurrence of dangerous situations (road work, hard-breaking, etc.). These services are supported by the exchange of broadcast CA and DEN messages (i.e. CAM and DENM) and are tailored-made for active safety and traffic efficiency applications. Connected automated vehicles (CAVs) can further improve their perception of the surrounding environment by exchanging sensor information (either raw data or processed data in the form of detected objects) with neighboring vehicles. The sharing of detected objects among CAVs will be performed by the Collective Perception Service (CPS) and will enable part of the Day2 applications and services [1]. The CPS is also under standardization at ETSI where its Technical Committee on ITS (Intelligent Transportation Systems) is defining, for instance, the format of the messages to be used for the exchange of the sensor information (known as Collective Perception Message - CPM), and their generation rules.

Most of the Day1 and Day2 applications will be supported by the broadcast exchange of V2X messages between vehicles (Vehicle-to-Vehicle, V2V), and between vehicles and the infrastructure (Vehicle-to-Infrastructure, V2I). All nearby vehicles and infrastructure nodes might be interested in the transmitted V2X messages and therefore most of them are transmitted in broadcast mode. In addition, broadcast transmissions do not require identifying the neighboring vehicles to contact them, which enables a fast exchange of information. This is particularly of interest in vehicular networks due to their dynamism and frequent topology changes. While unicast messages have their own mechanisms to assure their correct delivery, to date there are no such mechanisms for broadcast messages. Ensuring a high reliability in the delivery of V2X broadcast message is a challenge and a requirement of critical V2X applications/services. For example, consecutive failures in the reception of CPM messages from neighboring vehicles reduce the awareness of the driving environment that could provoke dangerous driving situations. In the event of a crossing pedestrian, certain vehicles approaching the pedestrian might not detect her/him using their built-in sensors. It would be of great value if these vehicles could be informed about the presence of this pedestrian. This could be performed by other vehicles that can detect the pedestrian and ensuring the delivery of the CPM messages. The motivation to design mechanisms that contribute to increasing the reliability of V2X broadcast messages has also raised recently due to the new 5.9 GHz band-plan in US [2]. This band-plan might result in that the 5.9 GHz band is not fully reserved for ITS services, and V2X and WiFi technologies would have to share it. In this case, the reliability of V2X broadcast transmissions might be compromised due to the increased channel usage.

To address these issues, current V2X communication technologies operating at 5.9 GHz are incorporating mechanisms to enhance the reliability of broadcast transmissions. For example, V2X communications based on cellular technologies (i.e. LTE-V and 5G NR) perform multiple transmissions of the same packet and exploit the HARQ (Hybrid Automatic Repeat Request) mechanism to combine

The work has been funded by the European Commission through the H2020 TransAID project (no. 723390), the Spanish Ministry of Economy, Industry, and Competitiveness, AEI, and FEDER funds (TEC2017-88612-R), the Spanish Ministry of Universities (IJC2018-036862-I).

Authors are with UWICORE laboratory, Universidad Miguel Hernandez de Elche, Elche (Alicante), Spain (e-mail: {bcoll, gthandavarayan, msepulcre, j.gozalvez}@umh.es).

the retransmissions and increase the likelihood to correctly receive the packet. IEEE 802.11bd, the amendment to the IEEE Std 802.11 standard operating at 5.9 GHz (a.k.a. IEEE 802.11p), is also being designed to support higher reliability for V2X broadcast transmissions. To this aim, IEEE 802.11bd is leveraging the MAC/PHY mechanisms that have been developed for the IEEE 802.11 technology during the past decade and that are part of the IEEE 802.11n, 802.11ac, 802.11ax amendments. At the PHY layer, this includes for example the support for midambles to achieve a better estimation of the channel and combat the Doppler effect, and advanced coding schemes such as LDPC (Low Density Parity Check). At the MAC layer, IEEE 802.11bd will support adaptive repetitions of V2X broadcast messages, with the number of repetitions varying based on the measured load of the V2X channel [3]. In addition, the use of acknowledgments has been recently proposed to assure the correct delivery of V2X broadcast messages. The proposed broadcast acknowledgment mechanism seeks controlling the load and signaling that might cause the feedback from all neighboring vehicles that receive a V2X broadcast message. To this aim, the proposed mechanism specifies that the V2X application should indicate the conditions/situations that trigger that the transmitting vehicle requests a feedback about the reception status of the V2X broadcast message to the receiving vehicles. The proposed mechanism also suggests that for controlling such load and signaling, the V2X application should also identify what receivers should reply to the broadcast acknowledgement request from the transmitting vehicle. These two decisions are out of the scope of the IEEE 802.11bd specification, and they should be defined based on the V2X application requirements.

Inspired by the discussions presented in IEEE 802.11bd, this paper proposes and evaluates a context-based broadcast acknowledgement mechanism that is used by the transmitting vehicles to selectively request the acknowledgment of specific (critical) broadcast messages. The proposed mechanism also implements retransmissions if the broadcast messages are not received correctly by the addressed receivers. The proposal has been evaluated for a CPS service that is designed to support the awareness of crossing pedestrians at intersections. The obtained results show that the proposed mechanism can contribute to increasing the reliability in the exchange of CPM messages between vehicles approaching to the intersection, and thus to reduce the potential risk that the vehicles approaching the intersection do not detect the pedestrians. In addition, the proposal is message and service agnostic, and could be used to increase the reliability of any other V2X broadcast message generated by any other applications or services.

II. CONTEXT-BASED BROADCAST ACKNOWLEDGEMENT

One of the major challenges for V2X technologies is to increase the reliability of V2X applications/services that are based on the (periodic) exchange of broadcast messages. Under the ongoing standardization framework of IEEE 802.11bd (a.k.a. Next Generation Vehicular networks, NGV), which is expected to finish by the end of 2021, it has been proposed the use of the context-based broadcast acknowledgement mechanism to assure the delivery of V2X broadcast messages. The design of this mechanism is being led by the company Autotalks which has presented in [4] and [5] different proposals to the IEEE P802.11-Task Group BD (NGV) [6]. Below it is described the mechanism that was presented by Autotalks in [7] that has been finally included in the motion booklet for IEEE 802.11bd and has inspired the proposal made in this paper.

*A. Concept*

The context-based broadcast acknowledgement mechanism presented in [7] aims at establishing a feedback loop between the transmitting vehicle and the receiver(s). The feedback loop is used as a report to inform the transmitter about the delivery status of the transmitted V2X broadcast messages. In IEEE 802.11 standards, this feedback loop is performed through the transmission of an acknowledgement (ACK) packet from the receiver to the transmitter. This acknowledgment mechanism has been used in the framework of IEEE 802.11 for unicast data transmissions. The acknowledgment mechanism for unicast has been used as the basis for broadcast data transmissions under the IEEE P802.11-Task Group BD (NGV).

Two important design aspects need to be taken into account in the context-based broadcast acknowledgement mechanism. The first one is related to the fact that multiple vehicles receiving a V2X broadcast message could send back ACK packets to the transmitting vehicle. These ACK packets would interfere with each other if they are not properly coordinated. If they interfere with each other, the transmitting vehicle would not properly receive any of them, which would trigger a retransmission of the broadcast message. To avoid this negative effect, ACK packets could be coordinated, but this is a complex task given the broadcast nature of the messages, and can significantly decrease the transmission efficiency. The second important design aspect is related to the network load and scalability. In general, the channel load and interference increase as the number of vehicles and the data traffic they generate augment. The increase of the interference augments the probability of packet loss and therefore the need to retransmit V2X messages. As a consequence, the interference is further increased due to the increase of message retransmissions. In Systems Theory, this is known as a positive feedback, defined as a process that amplifies changes and tends to move a system away from its equilibrium state and make it more unstable.

To avoid these situations, the context-based broadcast acknowledgment mechanism seeks controlling both: 1) what V2X broadcast messages need a feedback; and 2) the receiving vehicle(s) to whom a feedback should be requested. The mechanism proposed to the IEEE P802.11-Task Group BD does not specify any particular conditions or characteristics to identify any of the two. This is typically the case since IEEE P802.11 standards working groups focus on the specification, operation and protocols of lower layers of the OSI reference model (i.e. MAC/PHY), while these two decisions are to be made by the higher layers (i.e. facilities/application) [7].

*B. Proposal*

Inspired by [7], this study shows in Fig. 1 a graph diagram to integrate the context-based broadcast acknowledgement mechanism in the ETSI ITS protocol stack [8]. The ETSI ITS protocol stack focuses on the higher layers of the OSI reference model, that in ETSI ITS are referred to as Applications,

Facilities, and Networking & Transport. The ETSI ITS's Access layer integrates the IEEE 802.11 MAC/PHY layers.

At the higher layers (i.e. Application and Facilities layers), the 'V2X App' generates V2X broadcast messages[1]. Upon the generation of each of these messages, the 'V2X App' also checks whether the conditions to request an ACK packet from the receiver for this message are met (see block 'Conditions to request ACK' in Fig. 1). As indicated above, these conditions are specific and defined by the 'V2X App', e.g. based on reliability requirements. In case the conditions to request an ACK are met, the 'V2X App' creates a tag that is used to transfer information between layers within the V2X protocol stack and that is attached to the V2X broadcast message. This tag includes the unique ID of the V2X broadcast message that needs to be acknowledged (*V2XpktId*), and the ID of the receiver that will be requested to reply with the ACK packet (*RxId*). Then, the V2X broadcast message, together with the created tag, if the conditions are met, is passed down to the lower layers.

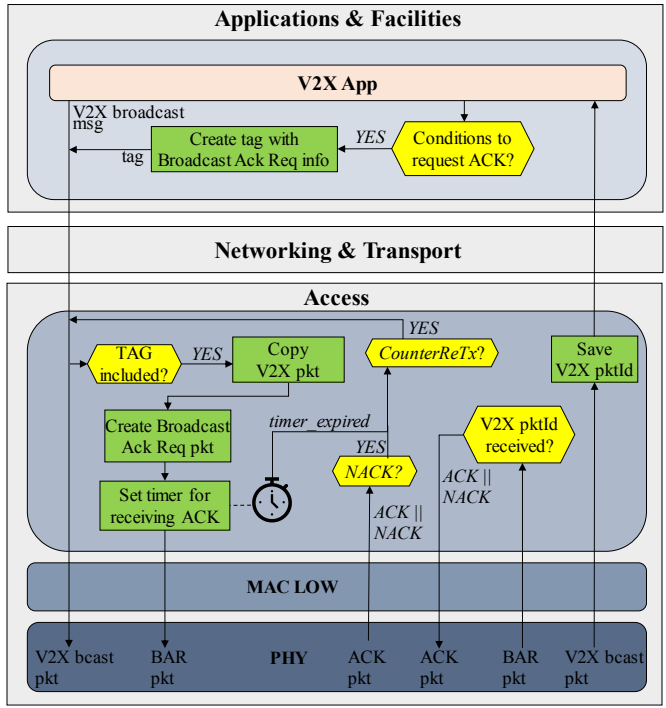

Fig. 1. Integration of the broadcast acknowledgment mechanism in the ETSI ITS protocol stack

When the V2X broadcast packet reaches the MAC layer, it is checked whether it has attached the tag (see 'TAG included?' in Fig. 1). In any case, the V2X broadcast packet is passed down to the lower layers for its transmission. In case the V2X broadcast message has attached the tag, a copy of the V2X packet is also saved at the MAC layer (see 'Copy V2X pkt' in Fig. 1). Besides, the MAC layer creates a Broadcast ACK Request (BAR) packet. This BAR packet is addressed to the receiver *RxId* identified by the V2X App. It should be noted the BAR packet is unicast and that the transmitter waits a limited time for the reception of the ACK from the addressed receiver (see 'Set timer for receiving ACK' in Fig. 1). Once the BAR packet is populated, it is passed down to the lower layers. If the transmitting vehicle receives the ACK packet, it checks whether the ACK packet shows that the addressed *RxId* vehicle received the V2X broadcast packet with ID *V2XpktId* (*ACK*) or not (*NACK*). It might also happen that the timer set by the transmitting vehicle to wait for the ACK packet expires before receiving any ACK packet. In case the ACK packet indicates *NACK* or the timer to wait for the ACK packet had expired, the transmitting vehicle retransmits the saved copy of the V2X broadcast packet. The number of retransmissions is limited to *CounterReTx*. The retransmitted V2X broadcast packet also triggers the transmission of the BAR packet, following similar procedures than the original transmission of the V2X broadcast packet. In case there are no remaining retransmissions, or the ACK packet indicates *ACK*, the MAC layer could send a report to the 'V2X App' that shows the result of the context-based broadcast acknowledgment mechanism.

[1] Note that information generated at the App layer is included in messages, while at bottom layers the information is included in packets.

Fig. 1 also shows the operation at the receiving vehicles. Upon the reception of the V2X broadcast packet, the MAC layer saves the ID of the received packet (*V2XpktId*). This saved *V2XpktId* is used, in the event that a BAR packet is received, to report back using the ACK packet whether a particular V2X broadcast packet was received (*ACK*) or not (*NACK*).

Fig. 2 shows the timing and sequence in the transmission of each of the packets. It should be noted that the BAR packet follows the V2X broadcast packet in accessing the medium and that there is no specific timing between them as it is shown in Fig. 2, i.e. they both follow the MAC back-off procedures to access the medium that are specified at the MAC LOW layer (see Fig. 1). On the other hand, the time between the BAR packet and the ACK packet transmitted by the addressed *RxId* vehicle is SIFS (Short Inter-Frame Space) as depicted in Fig. 2. Note that this is the regular operation at the MAC between unicast and ACK packets.

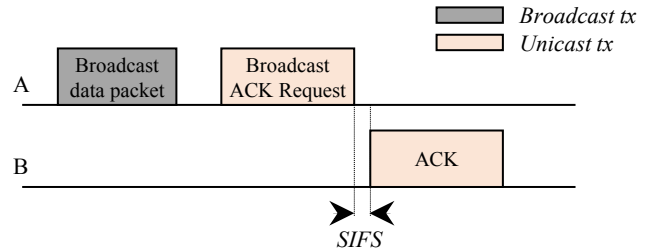

Fig. 2. Sequence and timing between the broadcast data packet, the broadcast ACK request packet and the ACK packet.

III. SCENARIO OF EVALUATION

*A. Simulation environment*

This work considers the urban intersection scenario depicted in Fig. 3. In the scenario there are vehicles stopped at the traffic light on the horizontal lane, and vehicles approaching to the intersection on the vertical lane. In the scenario, there is a potential risk that the vehicles approaching to the intersection from the vertical lane (e.g. vehicle B) do not detect the pedestrians that are crossing the street using their in-built sensors; e.g. LIDAR sensor that is typically used for pedestrians detection. In the considered scenario, pedestrians crossing the street are only detected by the vehicles that are stopped at the traffic light (e.g. vehicle A). Typically, it would be the first vehicle (vehicle A) since all other vehicles stopped behind would have their front sensors shadowed by the vehicles in front. To improve the awareness of the driving environment, the vehicles in the scenario implement the CPS service over the V2X radio interface. Based on the CPS service, vehicles frequently transmit CPM broadcast messages that include the objects detected by their built-in sensors; the details about the content and the generation of the CPM messages are reported in Section III.B. Detected objects might be either vehicles or pedestrians. Detected pedestrians are represented in Fig. 3 by red circles with a bounding box.

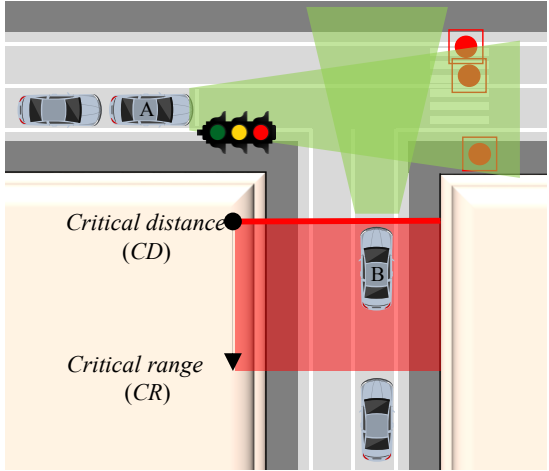

Fig. 3. Urban intersection scenario

The vehicles in the scenario utilize the context-based broadcast ACK mechanism introduced in Section II for the CPM messages. In particular, vehicles stopped at the traffic light transmit BAR messages to the vehicles driving on the vertical lane when one of the objects included in their CPM is a pedestrian. In order to identify the vehicle that is addressed in the BAR message, this study has defined a critical distance (*CD*) following the study presented in [9]. The *CD* is defined as the minimum distance to the intersection at which vehicles approaching the intersection on the vertical lane should receive a CPM that includes the pedestrian to avoid a potential collision to the pedestrians at the intersection. Considering a uniform deceleration model, as in [9], the critical distance can be computed as

$$CD = v \cdot RT + v^2/(2 \cdot a_{max}) \qquad (1)$$

where *v* represents the vehicle's speed, *RT* the (driver) reaction time, and $a_{max}$ the vehicles' emergency acceleration. Using the *CD* distance as a reference, the vehicle driving on the vertical lane that is closer to the *CD* distance is requested to acknowledge the BAR message. For the sake of scalability of the implemented broadcast acknowledgment mechanism, the proposed implementation limits the request of an ACK packet to the vehicles that are close to the intersection. In this context, this study has also defined a critical range (*CR*) from the *CD* distance to limit the search area of the vehicle that is requested to acknowledge the BAR message.

The studied scenario has been emulated in ns3. The simulator models radio signals propagation losses using the WINNER+ B1 model recommended by 3GPP in [10]. The model implements a log-distance pathloss model that differentiates between LOS and NLOS propagation conditions. It also models the shadowing using a log-normal random distribution with a standard deviation of 3dB (under LOS) and 4 (under NLOS). The vehicles are equipped with IEEE 802.11 radio access technology at 5.9 GHz, and transmit CPM messages with a data rate of 6 Mbps using the 1/2 quadrature phase shift keying (QPSK) transmission mode. The transmission power is set to 23 dBm (the antenna gain is 0 dBi). The vehicles driving on the vertical lane move at 20 m/s, and the traffic density is 50 vehicles/Km. Additional simulation parameters are summarized in TABLE I. .

TABLE I. SIMULATION PARAMETERS

| Variable | Value |
|---|---|
| **Scenario** | |
| Vehicles speed on the vertical lane (*v*) | 20m/s |
| Traffic density on the vertical lane | 50veh/Km |
| Critical Range (*CR*) | 40m |
| Reaction time (*RT*) | {0.75, 1, 1.25}s |
| Vehicles' emergency acceleration | 8m/s$^2$ |
| **V2X communications** | |
| Antenna gain | 0dBi |
| Data rate | 6Mbps (QPSK ½) |
| Noise figure | 9dB |
| Energy detection threshold | -85dBm |

### B. CPM's content and generation rules

The CPM messages are transmitted in the studied scenario following the ETSI CPS service's format and generation rules [11]. CPM messages include, among other, Perceived Object Containers (POCs) of 35 bytes each. POCs are optional and provide information about the detected objects (e.g. the distance between the detected object and the transmitting vehicle), the speed and dimensions of the object, and the time at which these measurements were taken.

CPM generation rules define how often a vehicle should generate a CPM and what information to be included in the CPM. Current ETSI CPM generation rules [11] states that a vehicle has to check every *T_GenCpm* if a new CPM should be generated and transmitted. For our analysis, *T_GenCpm* is set equal to the default 100ms. For every *T_GenCpm*, a vehicle should generate a new CPM if it has detected a new object. For previous detected objects, the vehicle should generate a CPM if any of the following conditions are satisfied:

a. For object class pedestrians (in general, Vulnerable Road Users - VRU) or animals:
  1. The last time any of the VRU or animal was included in a CPM was 0.5 (or more) seconds ago, include all VRU and animal in a CPM.

b. For object class different to VRU or animal:
  1. Its absolute position has changed by more than 4m since the last time its data was included in a CPM.
  2. Its absolute speed has changed by more than 0.5m/s since the last time its data was included in a CPM.

3. Its absolute velocity has changed by more than 4° since the last time its data was included in a CPM.
4. The last time the detected object was included in a CPM was 1 (or more) seconds ago.

A vehicle includes in a new CPM all new detected objects and those objects that satisfy at least one of the previous conditions. The vehicle still generates a CPM every 500ms (for objects that are categorized as VRU or animals) or every second (for other objects) even if none of the detected objects satisfy any of the previous conditions. In addition, the information about the onboard sensors is included in the CPM only once per second.

## IV. RESULTS

This section analyzes the performance of the proposed context-based broadcast acknowledgment mechanism in the scenario and under the conditions reported in Section III. The proposed mechanism, introduced in Section II, has been configured to utilize up to 3 retransmissions[2] (i.e. *CounterReTx*=3) when the transmitting vehicle does not receive the ACK packet. The evaluation also includes a reference technique as a baseline that follows the regular operation of the V2X CPS service, that does not implement the proposed context-based broadcast acknowledgment mechanisms, and that therefore does not perform retransmissions of the CPMs (i.e. *CounterReTx* = 0).

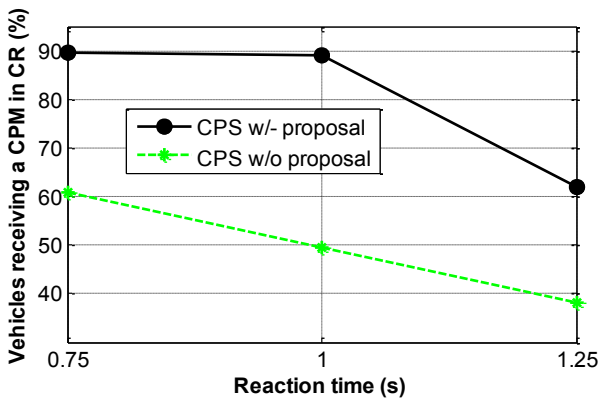

Fig. 4. Average percentage of vehicles that receive a CPM within the Critical Range (CR) as a function of the Reaction Time.

In the scenario under evaluation, it is of particular interest to analyze the percentage of vehicles approaching the intersection that receive at least one CPM, with information about the pedestrian, in the Critical Range. This percentage is shown in Fig. 4 as a function of the driver Reaction Time (*RT*), which results in a different location of the Critical Range using (1). Following [12], three different values have been considered in this study ranging from *RT*=0.75s to *RT*=1.25s. *RT*=1.25s can be considered a standard reaction time value [12]. This reaction time reduces in situations where the driver is aware that a dangerous situation might arise (e.g. approaching an intersection). Then, *RT*=1s and *RT*=0.75s have been also analyzed. The obtained results show that when *RT*=0.75s the regular V2X CPS service would successfully inform on average 60.8% of the vehicles driving on the vertical lane about the presence of the pedestrians crossing the street.

---

[2] This is the default limit of retransmissions for long packets in wireless networks.

As it is shown in Fig. 3, the vehicles driving on the vertical lane are not able to detect the crossing pedestrians using their built-in sensors. Therefore, ~40% of those vehicles are not aware of the presence of the crossing pedestrians with whom they could potentially collide. As the *RT* increases (i.e. the Critical Range moves away from the intersection), the percentage of vehicles that receive correctly a CPM decreases due to the worse propagation conditions. When the *RT*=1.25s, only 38% of the vehicles would be aware of the presence of the pedestrians crossing the street. Fig. 4 shows that when the vehicles implement the proposed context-based broadcast acknowledgment mechanism a higher percentage of vehicles receives at least a CPM within the Critical Range. This is due to the higher reliability in the delivery of the CPM messages. For example, when *RT*=1.25s, the percentage of vehicles that receives at least a CPM increases to 60.8%, and it reaches ~90% when the *RT*=0.75s.

Besides the analysis of the awareness of the crossing pedestrians in the Critical Range, this work has also studied the Object Awareness Ratio as a function of the distance between the object and the receiving vehicles. The Object Awareness Ratio is defined as the probability to detect an object (a crossing pedestrian in this study) through the reception of a CPM with its information in a given time window. Considering the CPM generation rules presented in Section III.B, a CPM with information about the crossing pedestrians is generated every 500ms. Therefore, the time window to calculate the Object Awareness Ratio is set to 500ms in this study. In this context, we consider that a crossing pedestrian is successfully detected by a vehicle if it receives at least one CPM with information about that pedestrian per 500ms. This analysis has been conducted considering *RT*=1s. In addition, different configuration of the proposed mechanism has been analyzed in order to assess the impact of the number of retransmissions in the reliability of the delivery of the CPM messages. In particular the proposed mechanism has been configured to utilize up to *CounterReTx*={1, 2, 3} retransmissions when the transmitting vehicle does not receive the ACK packet. The analysis is also conducted when the vehicles implement the regular V2X CPS service (i.e. *CounterReTx* = 0). Fig. 5 depicts the average Object Awareness Ratio as a function of the distance between the detected pedestrian and the vehicles receiving the CPM. In this study we focus on the vehicles approaching the intersection from the vertical lane. For example, the results reported in Fig. 5 show that when the CPS service does not implement the proposed mechanism, the probability that a vehicle that is 50m away to a crossing pedestrian detects that pedestrian through a received CPM is ~50%. This probability decreases when the distance between the vehicles receiving the CPM and the crossing pedestrian increases. The CPM messages with information about the crossing pedestrians are transmitted by the vehicles stopped at the traffic light that are under NLOS conditions to the vehicles approaching the intersection from the vertical lane (see Fig. 3). The distance between the crossing pedestrians and the vehicles approaching the intersection that need to receive the CPM message is similar to the distance between the vehicles stopped at the traffic light and the vehicles approaching the intersection. Therefore, the worse propagation effects with the increasing distances between the vehicles stopped at the traffic light and the vehicles approaching the intersection result in such reduction of the Object Awareness Ratio. The results reported in Fig. 5 show that when the vehicles implement the proposed mechanism, the probability to detect the crossing pedestrians increases. For example, when the proposed mechanism is configured to perform up to 3 retransmissions of the CPM (i.e. *CounterReTx*=3), the Object Awareness Ratio increases to ~82% when the distance between the vehicle and the crossing pedestrian is 50m. This significant increase demonstrates the potential of the proposed context-based broadcast acknowledgment mechanism to improve the reliability in the delivery of V2X broadcast messages when applied to CPMs.

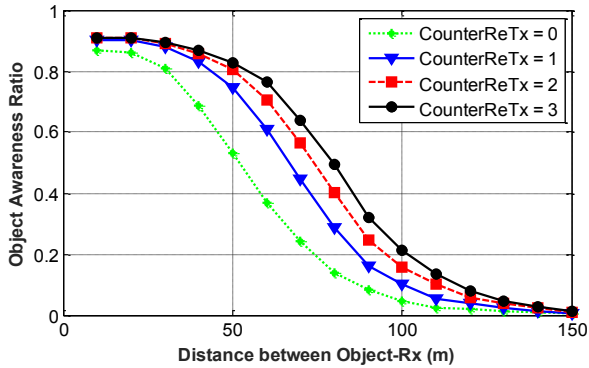

Fig. 5. Object Awareness Ratio as a function of the distance between the object (crossing pedestrian) and the vehicle receiving the CPM.

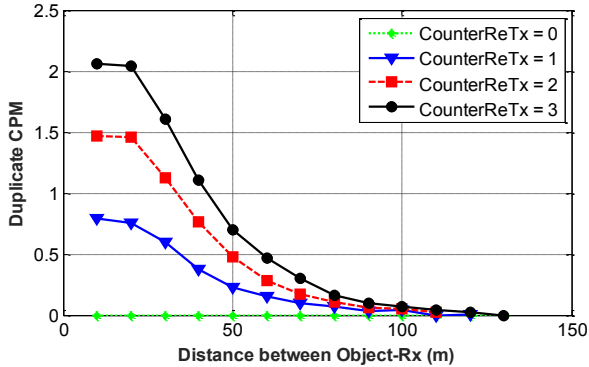

Fig. 6. Average received duplicate CPM messages as a function of the distance between the object (VRUs) and the vehicles receiving the CPM.

The proposed mechanism has shown to increase the Object Awareness Ratio (Fig. 5) thanks to the increased reliability in the delivery of the CPM messages. This is achieved by means of the context-based broadcast acknowledgment mechanism that performs a retransmission of the last transmitted CPM if the receiver indicates that the CPM was not received correctly via the ACK packet, or if the timer set to receive the ACK expires. However, it might also happen that the packet utilized to request the ACK (i.e. BAR packet), or the ACK itself, are not correctly received. If this is the case, the transmitting vehicle would perform a retransmission of a CPM that could have been correctly received, thereby generating a duplicate CPM message at the receiving vehicle. Fig. 6 depicts the average number of duplicate CPM messages that are received by the vehicles approaching the intersection, and that include information about the crossing pedestrians, as a function of the distance between the object and the receiver. As it can be observed in Fig. 6, the regular V2X CPS service (i.e. *CounterReTx* = 0) does not generate any duplicate CPM messages. However, with the proposed mechanism (i.e. *CounterReTx*={1, 2, 3}) there is certain probability that vehicles receive the same CPM multiple times. This effect is especially high at short distances, because all CPM transmissions and retransmissions are broadcast. Therefore, CPM retransmissions targeted to vehicles in the Critical Range are correctly received with high probability by vehicles that are close to the intersection (simply because they are closer and therefore propagation losses are lower). This side-effect is necessary to increase the probability of receiving a CPM in the Critical Range. The efficiency of the proposed mechanism at the system level is ensured by controlling the number of vehicles and situations where an ACK is requested.

## V. CONCLUSION

The reliability of V2X broadcast transmissions is challenged due to the lack of mechanisms to assure the correct delivery of messages. To improve the reliability of V2X broadcast transmissions, this paper proposes and evaluates a context-based broadcast acknowledgement mechanism. At the transmitting vehicle, the proposed mechanism selectively requests the acknowledgment of specific/critical V2X broadcast messages, and performs retransmissions at the MAC level if they are not correctly received. To improve the scalability of the proposed mechanism, the V2X applications/services are in charge of identifying the situations/conditions that trigger the execution of the broadcast acknowledgment mechanism, and the receiver(s) that should acknowledge the broadcast messages. While the proposed mechanism is valid for all types of V2X broadcast messages (e.g. CAMs, DENMs, CPMs, MCMs, etc.), this paper demonstrates its high potential considering the Collective Perception Service. The proposed mechanism is aligned with the discussions taking place at the IEEE P802.11-Task Group BD, and is therefore relevant for the standardization process of IEEE 802.11bd.